%% file: main.tex
\documentclass{article}
\usepackage{spconf,amsmath,graphicx, amsfonts, multirow, tabularx, booktabs, caption, float}
\usepackage{colortbl}
\usepackage{amssymb}
\usepackage{stmaryrd}
\usepackage[dvipsnames]{xcolor}
\definecolor{nicergreen}{rgb}{0.13, 0.54, 0.13}
\definecolor{nicered}{rgb}{0.83, 0.16, 0.16}
\definecolor{myhighlight}{rgb}{0.91, 0.95, 0.93}
\usepackage{hyperref}

\usepackage[fontsize=9pt]{fontsize}
\usepackage{bold-extra}
\usepackage[T1]{fontenc}

\usepackage{xspace}
\makeatletter
\def\onedot{\ifx\@let@token.\else.\null\fi\xspace}
\makeatother

\def\etal{\textit{et al}\onedot}
\def\eg{\textit{e.g}\onedot} 
\def\ie{\textit{i.e}\onedot} 
 
\def\etc{\textit{etc}\onedot} 
\newcommand{\app}{\raise.17ex\hbox{$\scriptstyle\sim$}}

\usepackage{tikz}
\newcounter{nodecount}

\usepackage{pifont}

\title{From Coarse to Fine: 
\\Efficient Training for Audio Spectrogram Transformers}
\name{Jiu Feng*, Mehmet Hamza Erol*, Joon Son Chung, Arda Senocak \thanks{*These authors contributed equally to this work. }}

\address{
Korea Advanced Institute of Science and Technology, South Korea}

\begin{document}
\maketitle

\input{sections/00_abstract}

\input{sections/01_intro}

\input{sections/02_approach}
\input{sections/03_experiments}

\input{sections/04_conclusion}

\vspace{-4mm}                                                                                                                                                                                                                                                                                                                                                                                                                                                                              
\section{Acknowledgment}
\vspace{-2mm}
This work was supported by the National Research Foundation of Korea grant funded by the Korean government (Ministry of Science and ICT, RS-2023-00212845) and the ITRC (Information Technology Research Center) support program (IITP-2024-RS-2023-00259991) supervised by the IITP (Institute for Information \& Communications Technology Planning \& Evaluation).

\newpage
\clearpage
\bibliographystyle{IEEEbib}
\bibliography{strings,refs}

\end{document}

%% file: sections/00_abstract.tex
\begin{abstract}
Transformers have become central to recent advances in audio classification. However, training an audio spectrogram transformer, \eg AST, from scratch can be resource and time-intensive. Furthermore, the complexity of transformers heavily depends on the input audio spectrogram size. In this work, we aim to optimize AST training by linking to the resolution in the time-axis. We introduce multi-phase training of audio spectrogram transformers by connecting the seminal idea of coarse-to-fine with transformer models. To achieve this, we propose a set of methods for temporal compression. By employing one of these methods, the transformer model learns from lower-resolution (coarse) data in the initial phases, and then is fine-tuned with high-resolution data later in a curriculum learning strategy. Experimental results demonstrate that the proposed training mechanism for AST leads to improved (or on-par) performance with faster convergence, \ie requiring fewer computational resources and less time. This approach is also generalizable to other AST-based methods regardless of their learning paradigms.
\end{abstract}

\begin{keywords}
Audio Spectrogram Transformers, Audio Classification, Efficient Training, Temporal Redundancy
\end{keywords}

%% file: sections/01_intro.tex
\section{Introduction}
\label{sec:intro}

Convolutional Neural Networks (CNNs) and, more recently, transformers have made a significant impact on numerous computer vision and audio processing tasks. Among these tasks, audio classification is a central research topic that assigns labels to the given audio inputs. The existing transformer-based approaches employ a patch-based system for audio classification~\cite{gong21b_interspeech,gong2022ssast,chen2022hts,koutini2021efficient,baade2022mae,huangmasked,niizumi2022masked,zhu2023multiscale}, where the input spectrograms are divided into fixed-size patches to create tokens as input for the transformer backbone. With the paradigm shift toward transformer-based approaches, an emerging thread of work also aims to explore efficient ways of optimizing the complexity of transformers, as it increases quadratically with the input sequence length. Recent works aim to reduce the quadratic complexity to make transformers more efficient for audio processing applications. Koutini \etal~\cite{koutini2021efficient} propose a method called Patchout, which efficiently drops patches while concurrently disentangling the positional encodings of both time and frequency axes.  Later, masked auto-encoder approach is employed to reduce the number of tokens, either through reconstruction objectives~\cite{baade2022mae,huangmasked,niizumi2022masked} or direct prediction of representations~\cite{niizumi2023masked} for masked input patches. Unlike the approach of dropping the input patches, HTS-AT~\cite{chen2022hts} leverages a hierarchical shifted window approach known as the Swin Transformer~\cite{liu2021swin}, originally employed in the domain of vision. Subsequently, another hierarchical strategy, known as multi-scale transformers~\cite{zhu2023multiscale,liu2023mmvit}, is applied in the audio domain by hierarchically expanding the channels while reducing the spatial resolution in the model.
\input{tables/teaser}

As aforementioned, the existing transformer-based approaches in the audio domain take spectrograms as input. When extracting the spectrograms, varying resolutions in time result in a different number of tokens to train the transformer backbone. Our goal is to link the resolution in the time-axis to efficient training of Audio Spectrogram Transformers~\cite{gong21b_interspeech}. We posit that employing different resolutions in training audio classification models can be both intuitive and beneficial. Firstly, as discussed in~\cite{liu2022simple,liu2024learning}, spectrograms may exhibit temporal redundancy. Audio patterns can be uniformly continuous or periodic~\cite{chen2021audio,senocak2023event}. Thus, shortening the time dimension can eliminate the redundancy in the input spectrogram, resulting in a reduction in computational cost and time. Secondly, following a fundamental principle in vision on input resolution, models that learn from coarse to fine-grained data can achieve performance improvements due to the scale invariance of the representations. 

The use of reduced input signals for computational efficiency in audio classification has been investigated by~\cite{colangelo2021progressive,liu2022simple} in the context of CNNs only. Xubo \etal~\cite{liu2022simple} propose simple pooling methods, such as max pooling, average pooling, and \etc, to eliminate temporally redundant information and enhance efficiency. By training the model with temporally reduced audio input, it performs similarly to the baseline that uses the original input, as shown in Table 1. However, training the audio spectrogram transformer (AST) directly with the temporally reduced input results in a significant performance drop compared to the baseline. This highlights that training efficient ASTs with lower-resolution inputs requires a different approach. We conjecture that training ASTs in a curriculum learning fashion, starting from coarse to fine-grained data, can help mitigate this issue~\cite{irandoust2022training,li2023reclip}. As shown in Table 1, curriculum learning (fine-tuning with higher resolution data) achieves comparable performance while also reducing computational costs and time.

\definecolor{Pool}{rgb}{204,0,0}
\definecolor{Patch}{rgb}{0, 0, 204}
\definecolor{Fshift}{rgb}{0.949, 0.639 ,0.388}
\begin{figure*}[tp]
\centering
\includegraphics[width=\textwidth]{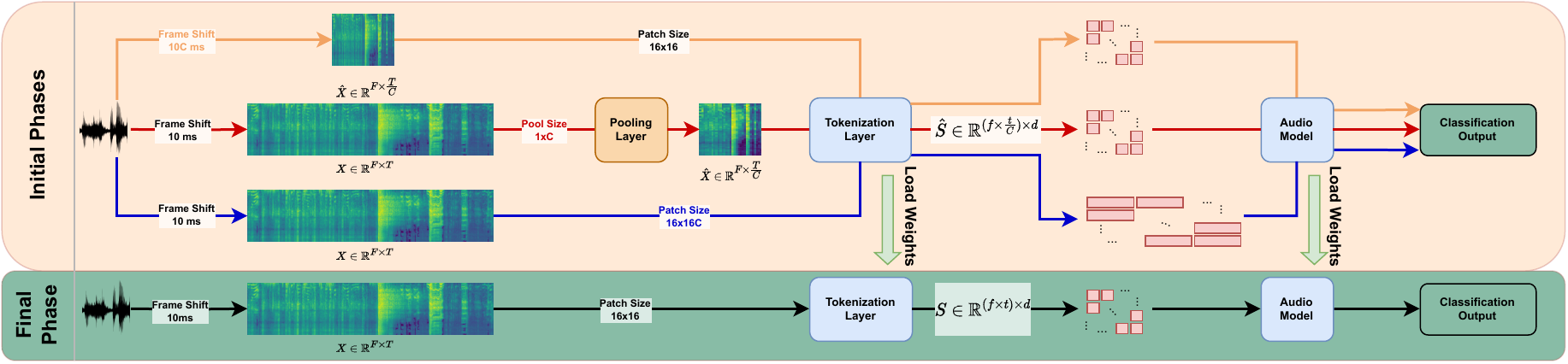} 
\caption{\textbf{Illustration of initial and final phase pipelines in our proposed training method.}~\textcolor{Fshift}{Fshift}, \textcolor{Pool}{Pool}, and \textcolor{Patch}{Patch} are compression methods from Section~\ref{compressMethods}. In the initial training phases, only one of them will be employed to get $f \times \frac{t}{C}$ number of tokens. Each method's unique contribution compared to the original pipeline is color-highlighted. Given numbers reflect the AST's original training settings.}
\vspace{-4mm}
\label{fig:pipeline} 
\end{figure*}

In this work, we propose curriculum learning-based training using the resolution of the audio signals as a proxy. This approach leads to efficient audio spectrogram transformer training, achieving comparable or better accuracy-to-computation tradeoffs compared to the widely used AST. To accomplish this task, we take the following steps (shown in Figure~\ref{fig:pipeline}):
(1) AST training is split into two (or multiple) phases. In the initial phase, the model is trained with input that has a lower resolution in the time-axis obtained through various reduction methods. Subsequent phases leverage higher resolution (eventually the original resolution) audio data. (2) We design various reduction methods, such as Frame-Shift, Pooling, and Patchification. One of these methods is employed during the initial phases of training. (3) While transitioning into subsequent phases, the model weights (\eg positional embeddings) are appropriately adapted to match the token numbers of the new phase. Based on these steps, our proposed training mechanism for AST~\cite{gong21b_interspeech} yields improved (or on-par) performance and requires fewer resources compared to the baseline model on four standard audio classification datasets. We conduct extensive ablation studies of our design choices. Moreover, we demonstrate that this approach can be further generalized to other AST-based approaches, such as HTS-AT~\cite{chen2022hts} and SSAST~\cite{gong2022ssast}.

%% file: tables/teaser.tex
\begin{table}[tp]
\centering
\scriptsize
\renewcommand{\tabularxcolumn}[1]{m{#1}}
\begin{tabularx}{\columnwidth}{ l >{\centering\arraybackslash}X >{\centering\arraybackslash}X >{\centering\arraybackslash}X}
\toprule
\multirow{1}{*}{\textbf{Method}} & \textbf{Architecture} & \textbf{Multi-Phase}  &\textbf{AudioSet (mAP)} \\
\toprule
Baseline - CNN14~\cite{kong2020panns} & CNN & \ding{55}  & 43.20 \\
Pool2~\cite{liu2022simple} & CNN & \ding{55}  & 42.60 \\
\midrule
Baseline - AST~\cite{gong21b_interspeech} & Transformer & \ding{55}  & 44.30 \\
Pool2 & Transformer & \ding{55}  & 42.79 \\
\rowcolor{myhighlight} Pool2$\shortrightarrow$1 & Transformer & \ding{51}  & 44.35 \\

\bottomrule
\end{tabularx}
\vspace{-2mm}
\caption{\textbf{Audio transformer models need multi-phase training for efficient learning.} Simple low-resolution training only works for CNNs. ``Pool2'' indicates training on 2 times compressed spectrograms. ``Pool$2\shortrightarrow1$'' denotes initial phase of training on 2 times compressed spectrograms, followed by training on full resolution.}
\vspace{-4mm}
\label{tab:teaser}
\end{table}

%% file: sections/02_approach.tex
\vspace{-4mm}
\section{Approach}
\label{sec:approach}

\vspace{-2mm}
\subsection{Mel-Spectrogram and Complexity}
\vspace{-2mm}

Given a waveform $W \in \mathbb{R}^{1 \times L}$, an Audio Spectrogram Transformer processes its corresponding mel-spectrogram (mel-spec), represented as $X=mel(W) \in \mathbb{R}^{F \times T}$, where $mel(\cdot)$ denotes the spectrogram generator module. This mel-spectrogram is first patchified and then tokenized into a sequence of tokens $S = Token(X) \in \mathbb{R}^{(f \times t) \times d}$, using the tokenization layer $Token(\cdot)$. The term $f \times t$ indicates the number of tokens. This sequence of tokens then serves as the input for the transformer's encoder layers. It is important to note that when using square-shaped patches, the length of the time axis has a significant impact on the token count. This, in turn, influences the complexity, which grows quadratically. Drawing from the insights of~\cite{liu2022simple}, we hypothesize that mel-specs may contain surplus temporal information during the early training phases, as the model might not need such a detailed representation initially. Therefore, by starting the training with a coarser temporal perspective and refining it progressively, we can improve the learning efficiency by potentially achieving comparable or better accuracy-to-computation tradeoffs compared to the traditional training methods.

\vspace{-2mm}
\subsection{Multi-phase Training}
\vspace{-2mm}

Our method splits the training into multiple phases. In the first phase, we apply one of the temporal compression methods (details are in Section~\ref{compressMethods}) to the input mel-specs, reducing the number of tokens along the time axis 
by a factor of $C$, yielding $ f \times \frac{t}{C}$ tokens. By introducing coarse data, the model can quickly assimilate generalized features, which guides the model weights into a good latent space. During the transition into the subsequent phases, the value of $C$ is reduced, and we transfer the trained weights from the previous phase by appropriately adapting the parameters that depend on the number of tokens (\eg interpolating the positional embeddings). Furthermore, the training settings, such as the learning rate, optimizer, and \etc, are reset to their initial values.

\vspace{-2mm}
\subsection{Compression Methods}\label{compressMethods}
\vspace{-2mm}
We investigate three distinct strategies for the temporal compression. Figure~\ref{fig:pipeline} provides an illustration depicting the roles of these methods.
\noindent \textbf{Change Frame-Shift Size (Fshift):} Mel-spectrograms are constructed by specifying frame-size and frame-shift values, which are, for instance, set to 25ms and 10ms by default in AST~\cite{gong21b_interspeech}. In this method, the frame-shift value is multiplied by a factor of $C$ when generating a mel-spec, which results in a temporally compressed mel-spec $\hat{X}$.

\vspace{-0.33cm}
\begin{equation}\label{eq:fshift}
     \hat{X} = mel'(W) \in \mathbb{R}^{F \times \frac{T}{C} }
\end{equation}

\noindent \textbf{Max/Avg Pooling (Pool):} Before tokenizing a mel-spec, we pass it through an additional max- or average-pooling layer with the kernel and stride of size $1 \times C$, resulting in a temporal reduction of the mel-spec by a factor of $C$.

\vspace{-0.33cm}
\begin{equation}\label{eq:pool}
\begin{aligned}
\text{Avg(X):}\;\ \hat{X} \left[i, j\right] &= \frac{1}{C}\sum_{n=0}^{C-1} X\left[i, C\cdot j+n\right] \\
\text{Max(X):}\;\ \hat{X} \left[i, j\right] &= \max _{n \in [0,C-1]} X\left[i, C\cdot j+n\right]
\end{aligned}
\end{equation}

\noindent where $i \in [0, F)$ and $j \in [0, \frac{T}{C})$.
\vspace{0.2cm}

\noindent \textbf{Flexible Patchification (Patch):} In the tokenization process, patches typically have a square shape, denoted as $p \times p$. Inspired by~\cite{feng2023flexiast}, we apply a rectangular patch size of $p \times Cp$ during tokenization. As each patch becomes $C$ times wider, the number of patches along the time dimension also decreases by a factor of $C$, resulting in temporal compression. Note that even though this method uses the full-resolution audio signal, we still consider it a compression method since it reduces the number of tokens. The models that employ this method apply either bilinear interpolation (BL) or PI-Resize operator (PI)~\cite{feng2023flexiast, beyer2023flexivit} when transferring the patch embedding weights from the previous phase. 

\vspace{-0.2cm}
\begin{equation}
\hat{S} = Token'(X) \in \mathbb{R}^{(f \times \frac{t}{C} ) \times d}\end{equation}

\input{tables/2phaseTimeSaving}
\input{tables/2phasePerformance}

%% file: tables/2phaseTimeSaving.tex
\begin{table*}[tp]
\centering
\scriptsize
\renewcommand{\tabularxcolumn}[1]{m{#1}}
\begin{tabularx}{\textwidth}{ l >{\centering\arraybackslash}c >{\centering\arraybackslash}X >{\centering\arraybackslash}X >{\centering\arraybackslash}c >{\centering\arraybackslash}X >{\centering\arraybackslash}X >{\centering\arraybackslash}c >{\centering\arraybackslash}X >{\centering\arraybackslash}X >{\centering\arraybackslash}c >{\centering\arraybackslash}X >{\centering\arraybackslash}X}
\toprule
\multirow{3}{*}{\textbf{Setting}} &  \multicolumn{3}{c}{\textbf{VGGSound}} & \multicolumn{3}{c}{\textbf{VoxCeleb}} & \multicolumn{3}{c}{\textbf{Kinetics-Sounds}} & \multicolumn{3}{c}{\textbf{AudioSet}}\\
\cmidrule(lr){2-4}
\cmidrule(lr){5-7}
\cmidrule(lr){8-10}
\cmidrule(lr){11-13}
& \textbf{Acc} & \textbf{FLOPs Save(\%)} & \textbf{Time Save(\%)} & \textbf{Acc} & \textbf{FLOPs Save(\%)} & \textbf{Time Save(\%)} & \textbf{Acc} & \textbf{FLOPs Save(\%)} & \textbf{Time Save(\%)} & \textbf{mAP} & \textbf{FLOPs Save(\%)} & \textbf{Time Save(\%)} \\
\toprule
Baseline & 49.40 & - & - & 41.90 & - & - & 62.92 & - & - & 44.30 & - & - \\
\midrule
Fshift 4$\shortrightarrow$1 & 49.90 {\tiny \textcolor{nicergreen}{(+0.50)}} & 58.31 & 55.00 & 41.95 {\tiny \textcolor{nicergreen}{(+0.05)}} & 42.56 & 38.89 & 63.84 {\tiny \textcolor{nicergreen}{(+0.92)}} & 42.56 & 38.89 & 43.82 {\tiny \textcolor{nicered}{(-0.48)}} & 35.32 & 33.40 \\
Fshift 2$\shortrightarrow$1 & 51.43 {\tiny \textcolor{nicergreen}{(+2.03)}} & 46.16 & 45.00 & 43.68 {\tiny \textcolor{nicergreen}{(+1.78)}} & 40.18 & 38.89 & 63.22 {\tiny \textcolor{nicergreen}{(+0.30)}} & 40.18 & 38.89 & 44.10 {\tiny \textcolor{nicered}{(-0.21)}} & 30.47 & 30.00 \\
\midrule
Avg Pool 4$\shortrightarrow$1 & 50.49 {\tiny \textcolor{nicergreen}{(+1.09)}} & 58.31 & 55.00 & 42.20 {\tiny \textcolor{nicergreen}{(+0.30)}} & 31.45 & 27.78 & 62.99 {\tiny \textcolor{nicergreen}{(+0.07)}} & 31.45 & 27.78 & 43.98 {\tiny \textcolor{nicered}{(-0.32)}} & 35.32 & 33.40 \\
Max Pool 4$\shortrightarrow$1 & 50.21 {\tiny \textcolor{nicergreen}{(+0.81)}} & 58.31 & 55.00 & 42.01 {\tiny \textcolor{nicergreen}{(+0.11)}} & 53.67 & 50.00 & 63.14 {\tiny \textcolor{nicergreen}{(+0.22)}} & 31.45 & 28.00 & 43.87 {\tiny \textcolor{nicered}{(-0.43)}} & 35.32 & 33.40 \\
Avg Pool 2$\shortrightarrow$1 & 49.85 {\tiny \textcolor{nicergreen}{(+0.45)}} & 46.16 & 45.00 & 43.75 {\tiny \textcolor{nicergreen}{(+1.85)}} & 40.18 & 38.89 & 63.14 {\tiny \textcolor{nicergreen}{(+0.22)}} & 29.07 & 27.78 & 43.98 {\tiny \textcolor{nicered}{(-0.32)}} & 30.47 & 30.00 \\
Max Pool 2$\shortrightarrow$1 & 49.66 {\tiny \textcolor{nicergreen}{(+0.26)}} & 46.16 & 45.00 & 43.95 {\tiny \textcolor{nicergreen}{(+2.05)}} & 40.18 & 38.89 & 63.66 {\tiny \textcolor{nicergreen}{(+0.74)}} & 17.96 & 16.67 & 44.19 {\tiny \textcolor{nicered}{(-0.11)}} & 30.47 & 30.00 \\
\midrule
Patch BL 4$\shortrightarrow$1 & 50.61 {\tiny \textcolor{nicergreen}{(+1.21)}} & 58.23 & 55.00 & 41.49 {\tiny \textcolor{nicered}{(-0.41)}} & 31.36 & 27.78 & 63.84 {\tiny \textcolor{nicergreen}{(+0.92)}} & 31.36 & 27.78 & 43.96 {\tiny \textcolor{nicered}{(-0.34)}} & 35.29 & 33.40 \\
Patch PI 4$\shortrightarrow$1 & 50.44 {\tiny \textcolor{nicergreen}{(+1.04)}} & 58.23 & 55.00 & 42.07 {\tiny \textcolor{nicergreen}{(+0.17)}} & 20.25 & 16.67 & 63.22 {\tiny \textcolor{nicergreen}{(+0.30)}} & 31.36 & 27.78 & 44.05 {\tiny \textcolor{nicered}{(-0.25)}} & 35.29 & 33.40 \\
Patch BL 2$\shortrightarrow$1 & 51.18 {\tiny \textcolor{nicergreen}{(+1.78)}} & 46.11 & 45.00 & 44.65 {\tiny \textcolor{nicergreen}{(+2.75)}} & 40.12 & 38.89 & 63.33 {\tiny \textcolor{nicergreen}{(+0.41)}} & 29.01 & 27.78 & 44.14 {\tiny \textcolor{nicered}{(-0.16)}} & 30.44 & 30.00 \\
Patch PI 2$\shortrightarrow$1 & 51.61 {\tiny \textcolor{nicergreen}{(+2.21)}} & 46.11 & 45.00 & 45.09 {\tiny \textcolor{nicergreen}{(+3.19)}} & 40.12 & 38.89 & 63.18 {\tiny \textcolor{nicergreen}{(+0.26)}} & 29.01 & 27.78 & 44.16 {\tiny \textcolor{nicered}{(-0.14)}} & 30.44 & 30.00 \\
\bottomrule
\end{tabularx}
\vspace{-2mm}
\caption{\textbf{FLOPs and time-saving ratios from 2-phase experiments using proposed compression methods.}}
\vspace{-4mm}
\label{tab:2phaseTimeSaving}
\end{table*}

%% file: tables/2phasePerformance.tex
\begin{table}[tp]
\centering
\scriptsize
\renewcommand{\tabularxcolumn}[1]{m{#1}}
\begin{tabularx}{\columnwidth}{ l >{\centering\arraybackslash}X >{\centering\arraybackslash}X >{\centering\arraybackslash}X >{\centering\arraybackslash}X}
\toprule
\textbf{Setting} & \textbf{VGGSound (Acc)} & \textbf{VoxCeleb (Acc)} & \textbf{Kinetics-Sounds (Acc)} & \textbf{AudioSet (mAP)} \\
\toprule
Baseline & 49.40 & 41.90 & 62.92 & 44.30 \\
\midrule
Fshift 4$\shortrightarrow$1 & 52.31 {\tiny \textcolor{nicergreen}{(+
2.91)}} & 43.22 {\tiny \textcolor{nicergreen}{(+1.32)}} & 64.14 {\tiny \textcolor{nicergreen}{(+1.22)}} & 44.17 {\tiny \textcolor{nicered}{(-0.13)}} \\
Fshift 2$\shortrightarrow$1 & 52.93 {\tiny \textcolor{nicergreen}{(+3.53)}} & 46.71 {\tiny \textcolor{nicergreen}{(+4.81)}} & 64.73 {\tiny \textcolor{nicergreen}{(+1.81)}} & 44.30 {\tiny \textcolor{Black}{(=)}} \\
\midrule
Avg Pool 4$\shortrightarrow$1 & 52.76 {\tiny \textcolor{nicergreen}{(+3.36)}} & 43.13 {\tiny \textcolor{nicergreen}{(+1.24)}} & 63.81 {\tiny \textcolor{nicergreen}{(+0.89)}} & 44.28 {\tiny \textcolor{nicered}{(-0.02)}} \\
Max Pool 4$\shortrightarrow$1 & 52.45 {\tiny \textcolor{nicergreen}{(+3.05)}} & 44.69 {\tiny \textcolor{nicergreen}{(+2.79)}} & 63.62 {\tiny \textcolor{nicergreen}{(+0.70)}} & 44.19 {\tiny \textcolor{nicered}{(-0.11)}} \\
Avg Pool 2$\shortrightarrow$1 & 53.42 {\tiny \textcolor{nicergreen}{(+4.02)}} & 46.47 {\tiny \textcolor{nicergreen}{(+4.57)}} & 64.21 {\tiny \textcolor{nicergreen}{(+1.29)}} & 44.22 {\tiny \textcolor{nicered}{(-0.08)}} \\
Max Pool 2$\shortrightarrow$1 & 53.17 {\tiny \textcolor{nicergreen}{(+3.77)}} & 46.64 {\tiny \textcolor{nicergreen}{(+4.74)}} & 64.03 {\tiny \textcolor{nicergreen}{(+1.11)}} & 44.35 {\tiny \textcolor{nicergreen}{(+0.05)}} \\
\midrule
Patch BL 4$\shortrightarrow$1 & 52.97 {\tiny \textcolor{nicergreen}{(+3.57)}} & 41.89 {\tiny \textcolor{nicered}{(-0.01)}} & 63.84 {\tiny \textcolor{nicergreen}{(+0.92)}} & 44.17 {\tiny \textcolor{nicered}{(-0.13)}} \\
Patch PI 4$\shortrightarrow$1 & 52.48 {\tiny \textcolor{nicergreen}{(+3.08)}} & 42.07 {\tiny \textcolor{nicergreen}{(+0.17)}} & 63.73 {\tiny \textcolor{nicergreen}{(+0.81)}} & 44.28 {\tiny \textcolor{nicered}{(-0.02)}} \\
Patch BL 2$\shortrightarrow$1 & 53.08 {\tiny \textcolor{nicergreen}{(+3.68)}} & 46.64 {\tiny \textcolor{nicergreen}{(+4.74)}} & 64.33 {\tiny \textcolor{nicergreen}{(+1.41)}} & 44.38 {\tiny \textcolor{nicergreen}{(+0.08)}} \\
Patch PI 2$\shortrightarrow$1 & 53.17 {\tiny \textcolor{nicergreen}{(+3.77)}} & 47.63 {\tiny \textcolor{nicergreen}{(+5.73)}} & 64.18 {\tiny \textcolor{nicergreen}{(+1.26)}} & 44.38 {\tiny \textcolor{nicergreen}{(+0.08)}} \\
\bottomrule

\end{tabularx}
\vspace{-2mm}
\captionof{table}{\textbf{Performance in the 2-phase approach when trained until convergence without training budget constraints.}}
\vspace{-4mm}
\label{tab:2phasePerformance}
\end{table}

%% file: sections/03_experiments.tex
\vspace{-4mm}

\section{Experiments}
\label{sec:experiments}
\vspace{-2mm}
\subsection{Datasets and Evaluation Metrics}\label{sec:dataset}
\vspace{-2mm}
\textbf{Datasets.} 
We conduct experiments using four datasets: (1) AudioSet, (2) VGGSound, (3) VoxCeleb, and (4) Kinetics-Sounds. AudioSet~\cite{gemmeke2017audio} is a large-scale multi-label dataset with approximately 2 million 10-second clips, featuring 527 labels across diverse audio categories. The balanced set is curated from the full set by selecting around 20K samples. VGGSound~\cite{VGGSound} consists of \app 200K 10-second videos labeled with 309 sound classes. VoxCeleb~\cite{nagrani2020voxceleb} provides an audio-visual dataset of human speech, containing 1251 speakers with approximately 145,000 utterances. Kinetics-Sounds is a subset of Kinetics~\cite{kay2017kinetics}, constructed from 10-second audio clips from YouTube. In our case, we use around 20K and 2.7K audio samples for training and testing, respectively.

\noindent \textbf{Evaluation metrics.} 
Due to the existence of multiple labels in each sample of AudioSet, we use mean average precision (mAP) across all classes for the evaluation. For the other datasets, we report the Top-1 classification accuracy (Acc) as samples are assigned only a single label.
\vspace{-2mm}
\subsection{Experiment setup}

\vspace{-2mm}
\textbf{Implementation details of baselines.} We train the AST baseline on AudioSet for 5 epochs by using the official configurations in~\cite{gong21b_interspeech}. 
For VGGSound, VoxCeleb, and Kinetics-Sounds, we train for 20 epochs and adopt the setup of the AudioSet training. However, mixup augmentation and weighted averaging are not utilized in these three datasets. To expedite the processing, non-overlapping patches are applied to all the datasets.
Note that, unlike ~\cite{gong2022ssast,baade2022mae}, our VoxCeleb experiments follow the same training and evaluation pipeline as the other three datasets, instead of the SUPERB framework~\cite{yang2021superb}, to maintain consistency in implementation and experiments. 
For the HTS-AT baseline on AudioSet, we adopt the full setting from ~\cite{chen2022hts} but train for 25 epochs and report the weighted averaging result of the top 15 checkpoints. On VGGSound, we train for 50 epochs but omit the weighted averaging result.
Lastly in SSAST, we follow the settings of~\cite{gong2022ssast} and perform self-supervised pretraining with patch-based model for 800k iterations on joint AudioSet and LibriSpeech~\cite{Panayotov2015Librispeech} datasets. The setups of supervised fine-tuning are identical to what we use in AST. More details are available at \href{https://sites.google.com/view/coarse-to-fine-audio}{https://sites.google.com/view/coarse-to-fine-audio}.

\input{tables/3phaseTimeSaving}
\input{tables/3phasePerformance}
\vspace{-2mm}
\subsection{Main Results}
\vspace{-2mm}
This section presents the results of our proposed training mechanism in terms of accuracy and resource efficiency, where training is split into two phases. We temporally compress the mel-spectrograms by a factor of $C$ during the initial phase, and we report the results for $C$ values of 2 and 4. The initial training phase is set to approximately 25$\%$ of the total number of training epochs in the baseline settings. Following this rule, our model is trained for 1 epoch on AudioSet and 5 epochs for the remaining datasets in the initial phase. Afterwards, we transfer the trained weights as the initialization of the final phase, where we use high-resolution (\ie original resolution) data for fine-tuning. When loading weights for the second training phase, we resize the positional embedding dimensions through bilinear interpolation to accommodate the change in the number of tokens.

\noindent \textbf{Accuracy/Computation trade-offs.} We analyze our model from the perspective of saving computational resources. To achieve this, we terminate the final phase of training at the earliest epoch that surpasses the baseline performance. All the differences in computational savings are calculated based on the baseline reference, where the epoch number with the highest accuracy is selected. The results presented in Table~\ref{tab:2phaseTimeSaving} show that we save from 18$\%$ to 58$\%$ of FLOPS while maintaining on-par or better performance than the baseline on different datasets. The only exception to this is AudioSet, where we save more than 30$\%$ of FLOPs with a negligible drop in mAP as in~\cite{liu2022simple}. Another observation is that there is no obvious difference between the compression methods. This highlights that the coarse-to-fine approach with time-axis compression is beneficial for the efficient training of AST, regardless of the compression method.

\noindent \textbf{Accuracy/No computational budget constraints.} In contrast to the previous analysis, here we allow the model to train until convergence without considering training budget constraints, solely aiming to achieve the highest performance improvement. As displayed in Table~\ref{tab:2phasePerformance}, AST consistently achieves further performance improvements, up to a 4$\%$ improvement on VGGSound, except for AudioSet where the improvements are negligible. These results demonstrate that the coarse-to-fine approach enables the model to begin learning from high-level information and gradually progress to important details, thus leading to better performance.

\vspace{-3mm}
\subsection{Ablation on Multi-Phases of Higher Resolution Fine-Tuning}
\vspace{-2mm}
In the main experiments, the model is trained with one low-resolution phase and one high-resolution fine-tuning phase. In this section, we further study the impact of using multiple phases for fine-tuning. Here, an additional phase is inserted between the initial phase and the final phase of fine-tuning, resulting in a three-phase training. Specifically, the model is sequentially trained with the (4$\shortrightarrow$2$\shortrightarrow$1) variant, where the model is first trained with a compression rate of $C=4$, and then progressively fine-tuned with $C=2$ and $C=1$, which represents the original resolution in the final phase. We schedule the intervals for the initial phases as 30$\%$ of the baseline total training, resulting in 3 epochs for the initial phases.

Similar to the main experiments, the termination of the final fine-tuning phase is decided based on two criteria: (1) the earliest epoch that surpasses the baseline performance, (2) training until convergence. While the first criterion is for exploring the Accuracy/Computation trade-offs perspective, the latter one is used to achieve the highest performance improvement without considering computational resource constraints. The results are shown in Table~\ref{tab:3phaseTimeSaving} and Table~\ref{tab:3phasePerformance}. As the results demonstrate, three-phase training also brings both training efficiency and performance improvements compared to the baseline. However, we observe that two-phase training provides competitive performance to three-phase training. Therefore, we use two-phase training for simplicity.

\vspace{-4mm}
\subsection{Generalization on Different Baselines}
\vspace{-2mm}
To demonstrate the general applicability of the coarse-to-fine training approach with time resolution reduction to other methods, we conduct experiments with recent methods, SSAST~\cite{gong2022ssast} and HTS-AT~\cite{chen2022hts}, by simply applying our proposed training mechanism to them. All of these baselines are audio spectrogram transformer (AST) based methods. For simplicity, we employ the Fshift compression method with the two-phase approach (4$\shortrightarrow$1) and the final
fine-tuning phase is terminated when it surpasses the baseline performance. The models are evaluated on the VGGSound and AudioSet datasets in these experiments. The results are shown in Table~\ref{tab:htastSSAST}.

\noindent \textbf{HTS-AT.} Similar to the main experiments, here we set the duration of the initial phase training to approximately 25$\%$ of the total number of training epochs in the HTS-AT baseline settings. As Table~\ref{tab:htastSSAST} illustrates, the coarse-to-fine training approach saves around 20$\%$ of training FLOPs while providing on-par performance with the baseline. Note that HTS-AT is already a very efficient transformer-based method, and our training paradigm further enhances its efficiency.

\noindent \textbf{SSAST.} Since SSAST is a self-supervised method, it undergoes pre-training before being applied to downstream tasks~\cite{gong2022ssast}. Generally, the pre-training phase is the most time and computation-intensive stage. Therefore, we apply our coarse-to-fine training paradigm during the pre-training stage. The model is initially pre-trained for 100K iterations with low-resolution data, followed by 500K iterations in the final phase with the original resolution. We utilize the joint AudioSet and LibriSpeech datasets for training, adhering to the baseline setting. As shown in Table~\ref{tab:htastSSAST}, the coarse-to-fine training paradigm leads to improved performance in downstream tasks, while simultaneously achieving more than a 30$\%$ reduction in training FLOPs and time. Moreover, these results indicate that the proposed training mechanism is generalizable to other AST-based methods, regardless of their learning paradigms, \ie whether supervised or self-supervised.
\input{tables/htastSSAST}

%% file: tables/3phaseTimeSaving.tex
\begin{table*}[tp]
\centering
\scriptsize
\renewcommand{\tabularxcolumn}[1]{m{#1}}
\begin{tabularx}{\textwidth}{ l >{\centering\arraybackslash}c >{\centering\arraybackslash}c >{\centering\arraybackslash}X >{\centering\arraybackslash}c >{\centering\arraybackslash}c >{\centering\arraybackslash}X >{\centering\arraybackslash}c >{\centering\arraybackslash}c >{\centering\arraybackslash}X}
\toprule
\multirow{2}{*}{\textbf{Setting}} & \multicolumn{3}{c}{\textbf{VGGSound}} & \multicolumn{3}{c}{\textbf{VoxCeleb}} & \multicolumn{3}{c}{\textbf{Kinetics-Sounds}} \\
\cmidrule(lr){2-4}
\cmidrule(lr){5-7}
\cmidrule(lr){8-10}
& \textbf{Acc} & \textbf{FLOPs Save (\%)} & \textbf{Time Save (\%)} & \textbf{Acc} & \textbf{FLOPs Save (\%)} & \textbf{Time Save (\%)} & \textbf{Acc} & \textbf{FLOPs Save (\%)} & \textbf{Time Save (\%)} \\
\toprule
Baseline & 49.40 & - & - & 41.90 & - & - & 62.92 & - & - \\
\midrule
Fshift 4$\shortrightarrow$2$\shortrightarrow$1 & 49.97 {\tiny \textcolor{nicergreen}{(+0.57)}} &58.68 & 56.00 & 42.83 {\tiny \textcolor{nicergreen}{(+0.93)}} & 31.87 & 28.89 & 64.36 {\tiny \textcolor{nicergreen}{(+1.44)}} & 42.98 & 40.00 \\
Avg Pool 4$\shortrightarrow$2$\shortrightarrow$1 & 50.54 {\tiny \textcolor{nicergreen}{(+1.14)}} & 58.68 & 56.00 & 42.42 {\tiny \textcolor{nicergreen}{(+0.52)}} & 42.98 & 40.00 & 63.44 {\tiny \textcolor{nicergreen}{(+0.52)}} & 42.98 & 40.00 \\
Max Pool 4$\shortrightarrow$2$\shortrightarrow$1 & 50.76 {\tiny \textcolor{nicergreen}{(+1.36)}} & 58.68 & 56.00 & 42.29 {\tiny \textcolor{nicergreen}{(+0.39)}} & 42.98 & 40.00 & 64.07 {\tiny \textcolor{nicergreen}{(+1.15)}} & 31.87 & 28.89 \\
Patch BL 4$\shortrightarrow$2$\shortrightarrow$1 & 50.09 {\tiny \textcolor{nicergreen}{(+0.69)}} & 58.60 & 56.00 & 41.29 {\tiny \textcolor{nicered}{(-0.61)}} & 20.67 & 17.78 & 63.99 {\tiny \textcolor{nicergreen}{(+1.07)}} & 42.89 & 40.00 \\
Patch PI 4$\shortrightarrow$2$\shortrightarrow$1 & 50.24 {\tiny \textcolor{nicergreen}{(+0.84)}} & 58.60 & 56.00 & 42.25 {\tiny \textcolor{nicergreen}{(+0.35)}} & 31.78 & 28.89 & 63.99 {\tiny \textcolor{nicergreen}{(+1.07)}} & 42.89 & 40.00 \\

\bottomrule
\end{tabularx}
\vspace{-2mm}
\caption{\textbf{FLOPs and time-saving ratios from 3-phase experiments using proposed compression methods.}}
\vspace{-4mm}
\label{tab:3phaseTimeSaving}
\end{table*}

%% file: tables/3phasePerformance.tex
\begin{table}[tp]
\centering
\scriptsize
\renewcommand{\tabularxcolumn}[1]{m{#1}}
\begin{tabularx}{\columnwidth}{ l >{\centering\arraybackslash}X >{\centering\arraybackslash}X >{\centering\arraybackslash}X}
\toprule
\multirow{2}{*}{\textbf{Setting}} & \textbf{VGGSound} & \textbf{VoxCeleb} & \textbf{Kinetics-Sounds} \\
& \textbf{(Acc)} & \textbf{(Acc)} & \textbf{(Acc)} \\
\toprule
Baseline & 49.40 & 41.90 & 62.92 \\
\midrule
Fshift 4$\shortrightarrow$2$\shortrightarrow$1 & 53.51 {\tiny \textcolor{nicergreen}{(+4.11)}} & 43.81 {\tiny \textcolor{nicergreen}{(+1.91)}} & 65.25 {\tiny \textcolor{nicergreen}{(+2.33)}} \\
Avg Pool 4$\shortrightarrow$2$\shortrightarrow$1 & 53.52 {\tiny \textcolor{nicergreen}{(+4.12)}} & 43.91 {\tiny \textcolor{nicergreen}{(+2.01)}} & 65.25 {\tiny \textcolor{nicergreen}{(+2.33)}} \\
Max Pool 4$\shortrightarrow$2$\shortrightarrow$1 & 53.48 {\tiny \textcolor{nicergreen}{(+4.08)}} & 44.41 {\tiny \textcolor{nicergreen}{(+2.51)}} & 64.58 {\tiny \textcolor{nicergreen}{(+1.66)}} \\
Patch BL 4$\shortrightarrow$2$\shortrightarrow$1 & 53.80 {\tiny \textcolor{nicergreen}{(+4.40)}} & 41.89 {\tiny \textcolor{nicered}{(-0.01)}} & 64.99 {\tiny \textcolor{nicergreen}{(+2.07)}} \\
Patch PI 4$\shortrightarrow$2$\shortrightarrow$1 & 53.62 {\tiny \textcolor{nicergreen}{(+4.22)}} & 43.11 {\tiny \textcolor{nicergreen}{(+1.21)}} & 65.43 {\tiny \textcolor{nicergreen}{(+2.51)}} \\

\bottomrule
\end{tabularx}
\vspace{-2mm}
\caption{\textbf{Performance in the 3-phase approach when trained until convergence.}}
\vspace{-4mm}
\label{tab:3phasePerformance}
\end{table}

%% file: tables/htastSSAST.tex
\begin{table}[tp]
\centering
\scriptsize
\renewcommand{\tabularxcolumn}[1]{m{#1}}
\begin{tabularx}{\columnwidth}{ l >{\centering\arraybackslash}X >{\centering\arraybackslash}c >{\centering\arraybackslash}X >{\centering\arraybackslash}c >{\centering\arraybackslash}X}
\toprule
\multirow{3}{*}{\textbf{Setting}} & \multirow{3}{*}{\parbox{1cm}{\centering \textbf{FLOPs Save (\%)}}} & \multicolumn{2}{c}{\textbf{AudioSet}} & \multicolumn{2}{c}{\textbf{VGGSound}} \\
\cmidrule(lr){3-4} \cmidrule(lr){5-6}
& & \textbf{mAP} & \textbf{Time Save (\%)} & \textbf{Acc} & \textbf{Time Save (\%)} \\
\toprule
Baseline HTS-AT & - & 46.92	& - & 52.82 & - \\
Fshift 4$\shortrightarrow$1 & 21.13 &  46.75 {\tiny \textcolor{nicered}{(-0.17)}}	& 10.50 & 52.98 {\tiny \textcolor{nicergreen}{(+0.16)}} & 14.32 \\
\midrule
Baseline SSAST &  - & 29.42	& - &  45.46 & - \\
Fshift 4$\shortrightarrow$1 & 34.58 &  30.28 {\tiny \textcolor{nicergreen}{(+0.87)}}	& 31.73 & 47.78 {\tiny \textcolor{nicergreen}{(+2.31)}}	& 31.73 \\
\bottomrule
\end{tabularx}
\vspace{-2mm}
\caption{\textbf{
Results for HTS-AT (using AudioSet-2M) and SSAST (using AudioSet-20K).} 
Experiments utilize the Fshift compression method in 2 phases, beginning with $C=4$. 
}
\vspace{-4mm}
\label{tab:htastSSAST}
\end{table}

%% file: sections/04_conclusion.tex
\vspace{-2mm}
\section{Conclusion}
\vspace{-2mm}
\label{sec:conclusion}
In this paper, we focus on the efficient training of audio spectrogram transformers with the motivation of temporal redundancy in spectrograms. We propose a coarse-to-fine training, initially using low-resolution input in the time-axis, progressively fine-tuning with higher resolution. Our experiments demonstrate on-par or better performance while saving computational resources. Furthermore, we show that this approach is generalizable to other AST-based methods. Transformers achieve optimal performance with large datasets~\cite{Dosovitskiy2021vit}. Considering that the audio domain does not have a dataset of similar size compared to vision yet, efficient training for audio transformers will play a crucial role in future research. Moreover, learnable schedulers for the phase transition can be also explored.